\documentclass[aps,epsf]{revtex4}   

\def\bra{\langle}
\def\ket{\rangle}

\begin{document}

\title{  Nuclear Incompressibility at Finite Temperature and Entropy}

\author{A.Z. Mekjian$^{a)}$, S.J. Lee$^{b)}$ and L. Zamick$^{a)}$}
                    
\address{$^{a)}$Department of Physics and Astronomy, Rutgers University
                             Piscataway, N.J. 08854}
\address{$^{b)}$Department of Physics and Institute of Natural Sciences,
                    Kyung Hee University, Suwon, KyungGiDo, Korea}



\begin{abstract}

  Features of the nuclear isothermal incompressibility $\kappa$
and adiabatic incompressibility $\kappa_Q$ are investigated.
The calculations are done at zero and finite temperatures
and non zero entropy and for several equations of state.
It is shown that $\kappa_Q$ decreases with increasing entropy
while the isothermal $\kappa$ increases with increasing $T$.
A duality is found between the adiabatic $\kappa_Q$ and the $T=0$
isothermal $\kappa$.
Our isothermal results are compared with a recent lattice Monte 
Carlo calculation done at finite $T$.
The necessity of including correlations is shown
if $\kappa$ is to have a peak with increasing $T$ as seen
in the Monte Carlo calculations.
A peak in $\kappa$ is linked to attractive scattering correlations
in two nucleons channel in the virial expansion in our approach
which are Pauli blocked at low $T$.

\end{abstract}

\pacs{ 
PACS no.: 21.65.+f, 21.30.Fe, 64.10.+h
 }

\maketitle
     

 The behavior of nuclear systems at moderately high temperature 
 and density is of current
 interest for several reasons. Such studies are important for understanding
 features of current medium energy collisions \cite{dasgup},
 for future RIA experiments, and for
 nuclear  astrophysics as in supernovae explosions.
 The incompressibility and its associated equation of state of pressure
 versus density and temperature is important in understanding flow
 produced in nuclear collisions as reviewed in Ref.\cite{daniel}
 and, in general, the thermodynamic properties of fermionic system
 under these conditions.
  This paper focuses on an 
 important quantity for understanding properties of these systems which is the
 nuclear incompressibility.  While the nuclear incompressibility at zero
 temperature has been studied for an extended period
 \cite{zamick,blaizot,pears},
 it is only relatively recently that its
 temperature dependence has been of concern.
 A recent investigation of M\"{u}ller etal \cite{muller}
 using quantum Monte Carlo techniques with nuclei on a lattice studied this
 temperature dependence and showed some interesting features about it.
 Namely, they have shown that the incompressibility coefficient can become very
 large at temperatures around 14 MeV. Large increases in the specific heat at
 constant volume was also seen in their quantum Monte Carlo approach.
 This increase in $C_V$ is also seen in Ref.\cite{dasmek}
 using a totally different approach based on recursive methods
 to obtain the finite temperature partition function of hadronic
 matter. In this later approach inhomogeneities  in the nuclear density are
 incorporated in the model in the form of clusters.  When a homogeneous system
 breaks into clusters at relatively low density,
 cost functions associated with large surface energies are 
 present and are responsible for the dramatic increase in $C_V$ with $T$.

 In this paper we will study the behavior of  the isothermal
 incompressibility with $T$ 
 and properties of the equation of state (EOS) at higher than normal density.
 As a baseline, we will begin with
 a mean field discussion of its behavior with $T$ to see how large the 
 incompressibility can become and if reaches the large values and associated
 peak found in the Monte-Carlo result of Ref.\cite{muller}.

 First, we define a quantity $\kappa$, the incompressibility coefficient, as
\begin{eqnarray}
   \kappa = k_F^2 \frac{d^2(E/A)}{d k_F^2} = 9 \rho^2 \frac{d^2(E/A)}{d\rho^2}
          = -9 V^2 \frac{d^2(E/A)}{dV^2}.     \label{kappadef}
\end{eqnarray}
 This quantity is evaluated at the saturation density $\rho_0$
 where $E/A$ has a minimum.
 The giant monopole resonance energy is then
 $E_0 = \sqrt{\frac{\hbar^2 \kappa A}{m \bra r^2\ket}}$.
 If the temperature is kept constant in the above derivatives we have the 
 isothermal incompressibility $\kappa$, and if the entropy is held fixed,
 the result is the adiabatic incompressibility $\kappa_Q$.
 The $\kappa$ and $\kappa_Q$ are equal at $T=0$ only.
 The quantity $\kappa$ defined above is not the isothermal compressibility
 defined in thermal physics as $K = -(\frac{1}{V}) ({dV}/{dP})_T$
 with $T$ held fixed and here $P$ is the pressure.
 Since $P=-dF/dV$, we
 have $K = (\frac{1}{V}) \left(1/\frac{d^2 F}{dV^2}\right)$. 
 At $T=0$, $F = E-T S = E$, and thus $\kappa = 9/(\rho_0 K)$.
 This reciprocal connection between $K$ and $\kappa$ is no longer true
 at finite $T$.  
 Besides the isothermal compressibility,
 an adiabatic compressibility $K_Q=-(\frac{1}{V}) (dV/dP)_S$ can be obtained
 by keeping the entropy constant.
 The reciprocal is related to $\kappa_Q$, the adiabatic incompressibility as
 $\kappa_Q=9/(\rho_0 K_Q)$.
 Since the natural variables for energy are entropy and volume from
 $dE=T dS- P dV$ variations of the energy with $V$ at constant $S$ bear
 a similar relation to variations of the  Helmholtz free energy
 with $V$ at constant $T$ where $dF=-S dT - P dV$.
 Thus the adiabatic incompressibility of Eq.(\ref{kappadef}) can go to zero
 for a Skyrme interaction as we shall see.
 The minimum point (also maximum point)  in the
 energy occurs at zero pressure since $dE/dV$ at constant entropy is $-P$.
 Therefore $E$ at constant $S$ has the same maximum and minimum points
 with variations in $V$ or $R$ or density as $F$ at constant $T$
 since both derivatives are $-P$ which is set to 0. 
The behavior of the adiabatic incompressibility is linked to a phase change.
As we shall see, the behavior of the isothermal incompressibility
may be associated with the appearance of a strongly corelated system
at high density and temperature.


    Our mean field discussion is based on a Skyrme interaction.
 To keep the discussion simple, we consider uncharged symmetric nuclear
 matter with no surface energy terms. The Skyrme interaction energy then is
  \begin{eqnarray}
            U/A= -a_0 \rho + a_\alpha \rho^{1+\alpha}.   \label{meanpot}
 \end{eqnarray}
 The $a_0$ term gives a medium range attraction while the $a_\alpha$ term
 is a short range repulsion.

 At $T=0$, the kinetic energy  $E_K/A = (\frac{3}{5}) E_F(\rho)$ with 
 the Fermi energy $E_F = \frac{\hbar^2}{2m} (\frac{6\pi^2}{4}\rho)^{2/3}$.
 The coefficients $a_0$
 and $a_\alpha$ are fixed to give a binding energy per particle $E_B/A=16$MeV
 at density $\rho = \rho_0 = 0.15$/fm$^3$, which gives
  $b_0 = a_0 \rho_0 =37 + 23/\alpha$ and 
  $b_\alpha = a_\alpha \rho_0^{1+\alpha} = 23/\alpha$ in MeV.
 The incompressibility coefficient $\kappa$ at $T=0$ is then 
  \begin{eqnarray}
    \kappa = - 2 E_K/A + 9 (1+\alpha) \alpha a_\alpha \rho_0^{1+\alpha}
         = 165 + 207 \alpha .
            \label{kappat0}
  \end{eqnarray}
For $\alpha=1/3$, $\kappa=234$MeV.
Smaller values of $\alpha$ lead to softer equations of state and lower $\kappa$.
 In the limit $\alpha \to 0$, logarithmic terms appear in Eq.(\ref{meanpot})
 coming from the presence of a factor
 $(x/\alpha) (1-x^\alpha) = \to -x\log(x)$. 
 The $x=\rho/\rho_0=(R_0/R)^3$ with $R_0^3=A/(\rho_0 4\pi/3)$.
 The $\alpha \to 0$ limit is the softest EOS allowed by Eq.(\ref{meanpot}),
 and this limit gives from Eq.(\ref{kappat0}) a value of $\kappa = 165$ MeV.
 A stiff EOS has $\alpha=1$ and $\kappa=372$ MeV. 
 Recent calculations done at $T=0$ \cite{piekar,soubb,vrete}
 have a value of $\kappa = 210 \sim 270$ MeV
and suggest a value of $\alpha=1/3$ in a Skyrme type approach.
A larger range of values of $\kappa$, from $211 \sim 350$ MeV,
were reported in Ref.\cite{lalazi}.
Because of these uncertainties in $\kappa$, from more realistic forces,
we will present results for various values of $\alpha$,
from $\alpha=0$ to $\alpha=1$.

 At non-zero $T \ll E_F$, the kinetic
 energy is $E_K/A = (\frac{3}{5}) E_F + (\frac{\pi^2}{4}) T^2/E_F$
\cite{huang}. 
 The energy per particle is
  \begin{eqnarray}
     E/A = 21 x^{2/3} + \left(\frac{\pi^2}{140}\right) \frac{T^2}{x^{2/3}}
               - b_0 x + b_\alpha x^{1+\alpha}.  \label{epa}
  \end{eqnarray}
 The value of $x$ that minimizes $E/A$ is $x_m$ and satisfies the equation:
 $14 (x_m^{2/3}-x_m^{1+\alpha}) - b_0 (x_m-x_m^{1+\alpha})
     = (\frac{\pi^2}{210}) T^2/x_m^{2/3}$.
Then the $\kappa = \kappa(T)$ is given by
  \begin{eqnarray}
    \kappa(T) = -42 x_m^{2/3} + 0.705 T^2/x_m^{2/3}
            + 9 \alpha (1+\alpha) b_\alpha x_m^{1+\alpha}  
  \end{eqnarray}
 At $T=2.5$, 5,  and 7.5MeV,  and for $\alpha=1/3$, the values of $\kappa$
 are 242, 265 and 302 MeV, respectively. 
 The corresponding values of $x_m$ are: 1.011, 1.043, 1.091.
 When $T$ is replaced with entropy per particle $S/A$ then this $T$ dependent
 term becomes
 $(E_F/\pi^2)(S/A)^2 = (35/\pi^2)(S/A)^2 x^{2/3}$
 since $S=(\pi^2/2) T A/E_F$ at low $T$.
 This $S/A$ term can simply be added to the 
 first term on the right side of Eq.(\ref{epa}) since both have the
 same $x^{2/3}$ dependence.

    If  the corrections to the nuclear matter incompressibility
 at $T=0$ from finite temperature terms are small,
 then these corrections can be obtained by using the following method.
 Let $E_0(R)$ be the nuclear matter energy per particle EOS and
 which has a minimum at $R_0$ and an incompressibility $\kappa_0$.
 If we add to this a term $E_x(R)$,
 so that $E(R)=E_0(R)+E_x(R)$ then the minimum shifts to a new point
 $R_m = R_0 + \Delta R_x$.
 The new minimum and $\kappa$ can be found by making a Taylor
 expansions around of $R_0$.
 The new $\kappa$ is
\begin{eqnarray} 
    \kappa = \kappa_0 + R^2\left(\frac{E_x}{A}\right)''
             - 2 R \left(\frac{E_x}{A}\right)' 
             - R\left(\frac{E_x}{A}\right)' 
                    \frac{Sk}{\kappa_0} 
                   \label{kappaex}
\end{eqnarray} 
 The various quantities are evaluated at $R_0$ and 
 each $'$ represents one derivative wrt $R$.
 Corrections to $\kappa$ involving the skewness $Sk=R^3 (E_0(R)/A)'''$
 or third derivative of the 
 energy were pointed out in Ref.\cite{blaizot,pears}.
Ellis et al \cite{ellis} used the correlation between 
compression modulus and skewness coefficient to examin
the implications in a relativistic Hartree Fock approximation
where the $E_x$ is the Coulomb interaction.
 The above expression is a modified  version of their result.
 Eq.(\ref{kappat0}) gives an expression for the incompressibility at $T=0$. 
 This will be $\kappa_0 = \kappa_0(\alpha)$ in Eq.(\ref{kappaex}).
 The skewness is $Sk(\alpha) = -3 (509 + 828 \alpha + 207 \alpha^2)$ in MeV.
 Comparing $Sk$ with $\kappa$ of Eq.(\ref{kappat0}) we see that the ratio
 of $Sk/\kappa_0$ is of the order of 10 and somewhat insensitive to $\alpha$.

 At low $T$, taking $E_x(R,T)= 0.0517 T^2 R^2/A^{2/3}$, we obtain the
   $\kappa(\alpha\to 0) = 165 + 1.16 T^2$,
   $\kappa(\alpha=1/3) = 234 + 1.32 T^2$ and
   $\kappa(\alpha=1) = 373 + 1.61 T^2$.
Thus we see that the first term is very sensitive to $\alpha$ but the finite
 temperature correction is somewhat insensitive to $\alpha$.
 At fixed entropy, the second derivative of $E(R)/A$ has a very different
 behavior than at fixed $T$. Namely, it decreases with $S/A$.
 This can easily be seen by noting that $E_x(R,S)=4.836(S/A)^2 A^{2/3}/R^2$
 compared to $E_x(R,T) =0.0517T^2 R^2/A^{2/3}$.
 We have the following final results (in MeV):
   $\kappa(\alpha\to 0) = 165 - 30 (S/A)^2$,
   $\kappa(\alpha=1/3) = 234 - 38 (S/A)^2$ and
   $\kappa(\alpha=1) = 373 - 53 (S/A)^2$.

    At higher $T$, the nearly degenerate Fermi gas kinetic energy term
 is replaced by a virial expansion in $\rho \lambda^3$,
 where $\lambda = \sqrt{\frac{2\pi\hbar}{mT}}$ is the quantum 
 wavelength.  Namely,  $E_K/A= (3/2) T (1+\sum_n c_n (\rho \lambda^3/4)^n)$
 with coefficients arising from antisymmetrization that are:
 $c_1=1/2^{5/2}=0.177$, $c_2=(1/8-2/3^{5/2})=-3.3 \times 10^{-3}$,
 $c_3= 1.11 \times 10^{-4}$, $\cdots$ \cite{huang,sjllg}.
 Since the $c_n$'s become small rapidly, we will keep terms up to $c_2$.
 Then $\kappa$ is given by:
  \begin{eqnarray}
    \kappa(T) = - 27 T (\lambda^3\rho_0/4)^2 c_2 x_m^2
       + 9 \alpha (1+\alpha) b_\alpha  x_m ^{1+\alpha}  \label{kappat} 
 \end{eqnarray}
 The $x_m$ is again the minimum of  $E/A$, but now evaluated with the new
 kinetic energy. The $x_m$ is affected by both the $c_1$ and $c_2$ terms
 at temperatures where $c_1$ dominants.
 A limiting value of $\kappa$ can be obtained by taking $T$ very large where
 $c_1$ term leads a minimum $x_m$ given by
 $x_m^\alpha=\frac{b_0}{(1+\alpha)b_\alpha}
   \left(1 - \frac{3}{2} T \frac{c_1}{b_0} \frac{\rho_0\lambda^3}{4}\right)$.
 In this high $T$ limit $\kappa$ is given by the second term on
 the right side of Eq.(\ref{kappat}) and is
\begin{eqnarray}
  \kappa(T\buildrel{>}\over{\sim} 10 {\rm MeV})
     = \kappa_{sat} \left(1 - \frac{3}{2} T \frac{c_1}{b_0}
            \frac{\rho_0\lambda^3}{4}\right)^{1+1/\alpha}
\end{eqnarray}
and goes to its saturation value 
$\kappa_{sat} = 9 \alpha b_0 (b_0/(1+\alpha) b_\alpha)^{1/\alpha}$
with a $T$ dependence of $1/\sqrt{T}$.
We note that the sign of $c_1$ determines whether it approaches from
above or below.
For purely antisymmetric correlations $c_1$ is positive because of the
statistical repulsion of fermions.
If $c_1$ becomes negative as will be discussed below 
it would approach from above. 
At infinite $T$ for $\alpha=1$, $\kappa = 704$MeV with $x_m=1.304$ or a minimum
density $\rho_m = 1.304 \rho_0$ and for $\alpha=1/3$
 $\kappa =  468$MeV with $x_m = 1.53$ or $\rho_m = 1.53 \rho_0$.
 In the limit $\alpha \to 0$, $\kappa = 380$ MeV with $x_m = e^{14/23} = 1.84$
 or $\rho_m = 1.84 \rho_0$.
 These are the limiting values for $\kappa$ and $x_m$.
The mean field results with antisymmetrization effects only
can be compared with the Monte Carlo results of Ref.\cite{muller}  
which show a very sharp increase in the $\kappa$  
with $T$ until a temperature of $\sim 14$ MeV is reached.
Then the incompressibility sharply decreases.
A peak in $\kappa$ of about 1500 MeV is present at a saturation density
of $2\rho_0 = 0.3$/fm$^3$.
From $T=20$ MeV to higher $T$'s the incompressibility is flat or saturates 
at a value of about 250 MeV.
It should also be noted that the saturating value of $\kappa$
obtained by Monte Carlo are lower than ours
even for the softest EOS with $\alpha \to 0$ limit.
The difference is due to finite size surface effect.
The lattice calculations are done for small systems.
An explanation of the maximum is qualitatively given in terms of clusters
that form which then repel each other through next nearest neighbor
interactions which is repulsive.
In our mean field model the incompressibility increases with $T$ 
and saturates at a value which is about twice its value at $T=0$.
The saturation density in our model also increases with $T$ and is somewhat
smaller than the Monte Carlo results of $2\rho_0$.

Before discussing how a peak in $\kappa$ may arise in our approach,
 we briefly investigate the case of constant entropy in the ideal gas limit
 using the Sackur-Tetrode law \cite{sakur}: $S/A=5/2-\ln(\lambda^3 \rho/4)$.
 This law connects $T$ to $\rho$ or $V$ as $T=C_S \rho^{2/3}$.
 Here $C_S = [2\pi(\hbar c)^2/(mc^2)] \exp[(2/3) (S/A)-5/3]$.
 The resulting $E(R)/A$ has a structure similar to the result for a degenerate 
 Fermi gas since both have a $\rho^{2/3}$ dependence for the kinetic energy
 term but with different coefficients.
 This feature and a similar result at lower $T$ suggests a duality 
 in the energy per particle EOS at constant entropy and its associated 
 $\kappa_Q$ and the $T=0$ EOS and its associated constant $T$ $\kappa$.   
We also note a parallel between $F$ as a function of $T$ and $V$ 
and $E$ as a function of $S$ and $V$.


We now turn to the issue of clusters or more precisely correlations
at moderately high $T$ and high $\rho > \rho_0$.
We study the corrections to the ideal gas law using the virial expansion
 $P = \rho T \left( 1 + c_1 (\rho\lambda^3/4) + c_2 (\rho\lambda^3/4)^2
      + \cdots \right)$.
If antisymmetry effects are the only corrections,
the coefficients can be calculated by following a procedure 
in Ref.\cite{chase1,chase2}
and are the coefficients already given before Eq.(\ref{kappat}).
This procedure is an extension of our fragmentation model by simply
noting that a cycle of length $k$ is analogous to a cluster of size $k$
with a weight function $x_k$.  
The grand canonical partition function is $\log Z = \sum x_k e^{\beta\mu k}$ 
and the mean number of cycles is $\bra n_k\ket = x_k e^{\beta\mu k}$ 
\cite{sjlnpa730}.
The pressure is $PV/T = \log Z = \sum x_k e^{\beta\mu k}$.
A constraint exists, $\bra A\ket = \sum k \bra n_k\ket = A$
which determines the fugacity $z = e^{\beta\mu}$  
in a power series in $A$ by inverting the series.
Then we arrive at
\begin{eqnarray}
 \frac{PV}{T} &=& A + \frac{-x_2}{x_1} A^2 
                 + \frac{4 x_2^2 - 2 x_1 x_3}{x_1^4} A^3
           \nonumber \\  & &
      + \frac{-20 x_2^3 + 18 x_1 x_2 x_3 - 3 x_1^2 x_4}{x_1^6} A^4
      + \cdots
\end{eqnarray}
Substituting $x_k = (-1)^{k+1} (V/\lambda^3)/k^{5/2}$ gives the desired 
power series in $(A/V) \lambda^3/4$ for fermions \cite{chase1,chase2}.
The factor of 4 is spin and isospin degeneracy.
The same procedure applies for bosons with $x_k = (V/\lambda^3)/k^{5/2}$.
For fragmentation in the Boltzmann limit 
the $x_k = \frac{V}{\lambda^3(k)} Z_{int}(k) = \frac{V}{\lambda^3(k)} e^{F_k/T}$
where $F_k$ is the internal free energy of a cluster of size $k$
and $\lambda(k) = \lambda/k^{1/2}$.
The effect of antisymmetry for odd $k$ clusters and symmetry for 
even $k$ clusters can be included.
The grand canonical ensemble represents a system of fermions (odd 
cluster sizes) obeying FD statistics and bosons (even cluster sizes)
obeying BE statistics.
The constraint of chemical equilibrium $\mu_k = k \mu_1$
or $\mu_k = z \mu_p + n \mu_n$ is imposed which determines the fugacity
from the constraint. In the $x_k$ model of
Ref.\cite{sjlnpa730,dasjen,mekjl64,sjlcan} this amounts to
 having various terms in $x_k$ that represent both cycles and clusters.
For example $x_1 = \frac{V}{\lambda^3}$, 
 $x_2 = - \frac{1}{2^{5/2}} \frac{V}{\lambda^3} 
       + 2^{3/2} \frac{V}{\lambda^3} Z_{int}(2)/4$ and
 $x_3 = \frac{1}{3^{5/2}} \frac{V}{\lambda^3}   
       + 3^{3/2} \frac{V}{\lambda^3} Z_{int}(3)/4$.
$x_4$ will have terms from the antisymmetry of monomers
from cycles of length 4, from symmetrization of dimers and from
clusters of size 4. 
Once the $x_k$'s are given the canonical partition function
can be generated by a recurrence relation \cite{sjlnpa730,chase3}.
A factor 1/4 appears from spin-isospin degeneracy which has been included.
The internal partiton $Z_{int}(k) = \sum g(E_j) e^{\beta E_j(k)}
 + \frac{1}{\pi} \sum_{J,T} \frac{(2J+1)(2T+1)}{\pi}
        \int \frac{d \delta_{J,T}}{d E} e^{-\beta E} dE$.
The sum is over bound state $E_j$ which have degeneracy $g(E_j)$ 
and $\delta_{J,T}$ is the phase shift in channel of spin $J$ and iospin $T$.
These phase shifts include effects from both attractive and repulsive
interactions.
Using nucleon-nucleon phase shifts the continuum contributions \cite{mekjc17}
reduces the bound state contribution by about 50\% for moderate
temperature ($T \sim 20$ MeV) and less for low temperatures
because of the Boltzmann weight factor in the integral.
At infinite $T$ $Z_{int} \to 0$ since the continuum exactly cancels
that bound states by Levinson's theorem \cite{mekjc17}.
As an initial example for $Z_{int}(2) 2^{3/2}$ we will 
consider $(1/2) (3/4) 2^{3/2} e^{|E_B|/T}$ to see how it compare with 
$1/2^{5/2}$;
1/2 is for the continuum reduction, 3 is for the spin degeneracy of the 
ground state of the deutron.
This choice has a value 1.06 neglecting $E_B$.
To reduce 1.06 to $1/2^{5/2}$ we would need a reduction factor or 1/6.
Also other spin isospin channels increase $Z_{int}$.
Thus $c_1$ can easily become minus.
For a negative $c_1$, the $\kappa$ is above its saturating value
and approaches it from above as $T^{-1/2}$.
At low $T$, $\kappa$ is below its saturating value and initially
increases as $T^2$ because the Fermi sea blocks excitations.
This behavior automatically implies a peak in $\kappa$.

Since we are at higher than normal density, we do not think of
real physical clusters,
but now view the presence of $k=2$ objects as correlated pairs
of fermions in various isospin and spin scattering channel
which are allowed since the Pauli blocking has been removed.
A deutron like structure may appear as a resonance and the
$e^{|E_B|/T} \to e^{-E_R/T}$ with $E_R$ the resonance energy.
Higher order $k=3$, 4, 5, ... terms represent higher order 
correlations of fermions in various $J$, $T$ channels.
For example $k=3$ can represent $J=1/2$, $T=1/2$ correlations.


In this paper we investigated the behavior of the infinite nuclear
matter incompressibility at finite temperature and entropy uing
a mean field theory
and also considering the role of correlations.
Various forms of the EOS are studied using a Skyrme parametrization.
Both the isothermal (constant temperature) and adiabatic (constant
entropy) incompressibilities are found
to be sensitive to the choice of the Skyrme repulsive parameter
$\alpha$ which gives the power of the density involved in
the repulsive term.
These two incompressibilities have very different behaviors.
The isothermal incompressibility increases with $T$ initially as $T^2$
until a saturation value is reached while the adiabatic incompressibility
decreases with increasing entropy and eventually goes to zero.
The isothermal incompressibility approaches its saturation value
as $- T^{-1/2}$ with the minus sign reflects an approach from below
the saturating value with increasing $T$.
This behavior arises from the statistical asymmetric repulsive 
correlations that represent the Pauli exclusion principle.
The adiabatic incompressibility is shown to arise from 
an equation of state (EOS) or energy per particle that has a
structure that is similar to a $T=0$ Fermi gas.

Our isothermal incompressibility results are compared with a 
recent Monte Carlo calculation
which gave a much stronger temperature dependence and much
larger value of the isothermal incompressibility.
The Monte Carlo result has a peak behavior in $\kappa$ 
with a peak of $\kappa = 1500$MeV at $T \sim 14$MeV.
The saturation point is at twice normal density in the
Monte Carlo result which is somewhat larger than the mean field 
saturation density.

We then discussed how a peak can appear in the isothermal 
incompressibility by looking at coefficents in the virial expansion, and 
in particular, we investigated the first coefficient called $c_1$ here.
The approach of $\kappa$ to its saturation value was shown to be
related to the sign of $c_1$, with $c_1$ positive having an approach 
from below and $c_1$ negative having an approach to the saturating 
value from above it.
The role of attractive correlation between nucleons was studied and 
shown to most likely be strong enough in various spin isospin channels
to overcome the statistical repulsion term at high $T$. 
The Pauli blocking of scattering is reduced at high $T$ and such 
correlations are present even a density $\rho > \rho_0$.
A two nucleon correlation of paired fermions in a high
density, but also high temperature, medium can account for the 
existence of a peak.
The structure of the peak is related to the presence of strongly
correlated fermions in triplets, and higher order correlations.

This work was supported 
in part by the DOE Grant No. DE-FG02-96ER-40987 and DE-FG02-95ER-40940 and
in part by Grant No. R05-2001-000-00097-0 from the Basic Research Program 
of the Korea Science and Engineering Foundation.

\end{document}